\documentclass[%
reprint,
showpacs,
 amsmath,amssymb,
 aps,
pra,
floatfix,
]{revtex4-2}

\usepackage{graphicx}

\begin{document}

\title{Unified Model for a Nonlinear Pulse Propagation in Composites and Optimization of THz Generation}

\author{A. Husakou}
\email{gusakov@mbi-berlin.de}
\affiliation{Max Born Institute, Max Born Str. 2a, 12489 Berlin, Germany}

\author{O. Fedotova}
\author{R. Rusetsky}
\affiliation{Scientific-Practical Materials Research Centre of   National Academy of Sciences of Belarus, P.Brovki str. 19, 220072 Minsk, Belarus}

\author{T. Smirnova}
\affiliation{Belarus State University, Niezaliežnasci ave 4, 220030 Minsk, Belarus}

\author{O. Khasanov}
\affiliation{Scientific-Practical Materials Research Centre of   National Academy of Sciences of Belarus, P.Brovki str. 19, 220072 Minsk, Belarus}

\author{A. Fedotov}
\affiliation{Belarus State University, Niezaliežnasci ave 4, 220030 Minsk, Belarus}

\author{T. Apostolova}
\affiliation{Institute for Nuclear Research and Nuclear Energy, Bulgarian Academy of Sciences, Tsarigradsko Chausse 72, 1784 Sofia, Bulgaria}
\affiliation{Institute for Advanced Physical Studies, New Bulgarian University, 1618 Sofia, Bulgaria}

\author{I. Babushkin}
\affiliation{Institute of Quantum Optics, Leibniz University Hannover, Welfengarten 1, 30167, Hannover, Germany}
\affiliation{Cluster of Excellence PhoenixD (Photonics, Optics, and Engineering-Innovation Across Disciplines), 30167, Hannover, Germany}
\affiliation{Max Born Institute, Max-Born-Str.~2a, 12489, Berlin, Germany}

\author{U. Sapaev}

\affiliation{Tashkent State Technical University, University street 2, 100097
Tashkent, Uzbekistan}

\begin{abstract}
We describe a unified numerical model which allows fast and accurate simulation of nonlinear light propagation in nanoparticle composites, including various effects such as group velocity dispersion, second- and third-order nonlinearity, quasi-free-carrier formation and plasma contribution, exciton dynamics, scattering and so on. A developed software package SOLPIC is made available for the community. Using this model, we analyze and optimize efficient generation of THz radiation by two-color pulses in ZnO/fused silica composite, predicting an efficiency of 3\%. We compare the role of various nonlinear effects contributing to the frequency conversion, and show that optimum conditions of THz generation differ from those expected intuitively.
\end{abstract}

\maketitle

\section{Introduction}
THz technology has attracted a lot of attention in the recent years, since it provides unique experimental tools and techniques in nonlinear and time-domain spectroscopy, biology and medicine, remote sensing,  security screening, as well as information and communication systems (see e.g. \cite{thz1,thz2,thz3,thz4}). For generation of THz radiation, different techniques were proposed, such as two-color ionizing femtosecond pulses in gases \cite{thz5,thz6,thz7,thz9,thz10,thz_our}, surface plasmas\cite{Herzer}, as well as optical rectification of intense ultrashort pulses in non-linear crystals \cite{cr1,cr2,cr3, cr4,hhgsi} which provide a basis for compact low-intensity devices. 
The needs of the THz technology require, however, extension of the range of the available techniques and materials, in order to provide a flexible design required in multifarious applications. Following this line, investigations of THz generation in various media such as water \cite{gen_xxx}, strongly magnetized plasma \cite{gen_yyy}, and centro-symmetric two-photon resonant molecular impurities \cite{two_q} were performed. Emission of terahertz radiation with broad bandwidth by femtosecond photoexcitation of spintronic materials (ferromagnetic and synthetic multiferroic heterostructures) was also reported recently \cite{Zhou,Agarwal}.

Nanoparticle (NP) composites were actively investigated in the past as a nonlinear material, e.g. \cite{comp_nl1,comp_nl2,comp_nl3}, and their particular strength lies in the flexibility of their design leading to unusual properties such as e.g. negative refractive index \cite{neg_n}. However, surprisingly, up to our knowledge they have not attracted attention as a medium for THz generation. In this paper, we close this gap by concepting a numerical model suitable for simulation of THz generation in nanocomposites. 
A range of linear and nonlinear effects such as group velocity dispersion, second- and third-order nonlinearity, quasi-free-carrier formation, exciton dynamics and so on are encompassed by the developed model. We use it to explore THz generation by two-color pulses in nanoparticle composites, to elucidate the contributions of different frequency conversion mechanisms, and to predict  efficiencies in few-percent range.  

The applicability of the above model is, in fact, much broader than mere simulation of THz generation; a wide range of nonlinear effects such as soliton dynamics and supercontinuum generation, frequency conversion, multi-level dynamics and electromagnetically-induced transparency, and so on can be studied using this unified approach. With this is mind, we have created the extensive documentation of the code and made the code publicly available \cite{code}, in a hope that it will be useful to the optical community for investigations of the nonlinear processes in nanocomposites and other materials.

The paper is organized as follows. In Section 2, we present the numerical model, including detailed formalism for all the relevant mechanisms. In Section 3, we optimize the THz generation by two-color pulses, and analyze the role of different parameters. A summary of the paper is given in the conclusion.

\section{Theoretical model}
We consider a composite consisting of two components, a homogeneous host material and spherical NPs (inclusions) randomly distributed in space. We assume a sufficiently low (typically few percent or below) filling fraction of the inclusions so that neither percolation nor interaction between the inclusions play a role. We consider homogeneous inclusions to be sufficiently small with diameter below the light wavelength so that effective-medium theory can be applied. Note that we do not place any limitations on the nature of host and inclusion materials, i.e., either of them could be a dielectric, a metal, or a semiconductor. We do not require point symmetry in host material or in the inclusions, so that the second-order susceptibility can be non-zero in either material. The model is designed to simulate light propagation over relatively short distances of few millimeters, below the damage threshold, and without back-reflection, therefore (1+1)D treatment using unidirectional propagation equation \cite{hh} is the most suitable.

Under these conditions, the following effects have to be taken into account: linear dispersion including intrinsic and scattering losses, second- and third-order optical nonlinearities and photoionization accompanied by ionization losses and plasma dynamics. In addition, transitions between excitonic states can play a significant role in the inclusion response, in particular for the generation of new frequencies in the THz range. Among the effects which were neglected in this treatment are thermal effects (due to slow ns-scale response), coupling to phonons (because of relatively slow ps-scale response), Raman scattering (which is typically weaker than instantaneous nonlinearities), anisotropy of the host material (due to the manufacturing limitations for composites), deviations of the inclusion from a sphere (because of typical manufacturing conditions), and generation of high-order harmonics (because of the considered intensity ranges).

The following unidirectional propagation equation is used to model the light propagation in a homogeneous medium \cite{hh,mg}:

\begin{eqnarray}
\frac{\partial E(z,\omega)}{\partial z}&=&-i\left(\frac{[\sqrt{\epsilon(\omega)}-n_g]\omega}{c}
-\beta(\omega_0)\right)E(z,\omega)\nonumber\\&-&\frac{i\omega}{2c\sqrt{\epsilon(\omega)}}P_{\mathrm{NL}}(z,\omega),
\label{main_e}
\end{eqnarray}

where $E(z,\omega)=\hat{F}E(z,t)=\int_{-\infty}^\infty E(z,t)\exp(-i\omega t) dt$ is the Fourier transform $\hat{F}$ of the electric field $E(z,t)$, $z$ is the propagation coordinate, $\epsilon(\omega)$ is the linear dielectric permittivity (generally speaking, complex-valued to include loss mechanisms), $n_g$ is the group refractive index, $\omega_0$ is a characteristic frequency of the pulse spectrum, $\beta(\omega)=\sqrt{\epsilon(\omega)}\omega/c$,  and $P_{\mathrm{NL}}(z,\omega)$ is the Fourier transform of the nonlinear part of the polarization. We would like to empathise that no slowly-varying envelope approximation was used, and $E(z,t)$ represents the real-valued field including the carrier oscillations. This approach yields a unified treatment for a pulse with arbitrary spectral content, which is particularly important for extremely broad spectra.

\subsection{Linear dispersion}The effective-medium theory allows to substitute the composite material by a homogenised medium with appropriately defined effective material parameters.  The effective refractive index of a composite can be expressed as \cite{mg}

\begin{eqnarray}
n_{\mathrm{eff}}&=&\left[(1-f)\epsilon_h+f\epsilon_i\frac{3\epsilon_h}{2\epsilon_h+\epsilon_i}\right.\nonumber\\&+&\left.2i\left(\frac{\epsilon_h-\epsilon_i}{2\epsilon_h+\epsilon_i}\right)^2\left(\frac{r_{\mathrm{NP}}\omega\sqrt{\epsilon_h}}{c}\right)^3\right]^{1/2},
\label{eps_eff}
\end{eqnarray}

where $f$ is the volume filling factor of the inclusions, $r_{\mathrm{NP}}$ is their radius, and $\epsilon_{h,i}$ are the frequency-dependent dielectric functions of the host and of the inclusions, correspondingly. The last term in the square brackets describes scattering losses.

\subsection{Second- and third-order nonlinearities}The second- and third-order nonlinear processes can also be described in the framework of the effective-medium theory. The expression for the effective second-order susceptibility looks like \cite{sb}

\begin{eqnarray}
    \chi_{\mathrm{eff}}^{(2)}(\omega_1&=&\omega_2+\omega_3;\omega_2,\omega_3)=(1-f)\chi_{h}^{(2)}+\nonumber\\&+&fx(\omega_1)x(\omega_2)x(\omega_3)\chi_{i}^{(2)},
\end{eqnarray}

where $\chi_{h}^{(2)}$ and $\chi_{i}^{(2)}$ are the susceptibilities of host and inclusion materials, correspondingly. Note that we neglected the frequency dependence of the susceptibilities of host and inclusions, which is a good assumption far from resonances. Quantity $x(\omega)$ is the ratio of local field inside the inclusion and the incident field:

\begin{equation}
    x(\omega)=\frac{3\epsilon_h(\omega)}{2\epsilon_h(\omega)+\epsilon_i(\omega)}.
\end{equation}

Here we note that, due to photoionization as described below, the $\epsilon_i(\omega)$ and therefore $x$, strictly speaking, depend on time due to buildup of plasma during the pulse. However, in the  current simulation we neglect this dependence, assuming that corresponding change of $\epsilon_i(\omega)$ is small and that we are far from the plasmonic resonance given by $2\epsilon_h(\omega)=-\epsilon_i(\omega)$.

Similarly, for the effective third-order susceptibility we write \cite{sb}

\begin{eqnarray}
    \chi_{\mathrm{eff}}^{(3)}(\omega_1&=&\omega_2+\omega_3+\omega_4;\omega_2,\omega_3,\omega_4)=(1-f)\chi_{h}^{(3)}+\nonumber\\&+&fx(\omega_1)x(\omega_2)x(\omega_3)x(\omega_4)\chi_{i}^{(3)},
\end{eqnarray}

where $\chi_{h}^{(3)}$ and $\chi_{i}^{(3)}$ are the susceptibilities of host and inclusion materials, correspondingly. The final expressions which were used to calculate the corresponding polarizations look like

\begin{eqnarray}
    P_{\chi^{(2)}}(z,\omega)&=&(1-f)\epsilon_0\chi_{h}^{(2)}\hat{F}E(z,t)^2+\nonumber\\&+&f\epsilon_0\chi_{i}^{(2)}\hat{F}[\hat{F}^{-1}\{E(z,\omega)x(\omega)\}^2],
    \end{eqnarray}
    
\begin{eqnarray}
    P_{\chi^{(3)}}(z,\omega)&=&(1-f)\epsilon_0\chi_{h}^{(3)}\hat{F}E(z,t)^3+\nonumber\\&+&f\epsilon_0\chi_{i}^{(3)}\hat{F}[\hat{F}^{-1}\{E(z,\omega)x(\omega)\}^3].
\end{eqnarray} 

\subsection{Plasma dynamics}Let us turn to the description of plasma formation and dynamics. In the framework of SOLPIC, we consider a case when the ionization potential $I_p$ of the inclusions is lower than that of the host material, so that due to the sensitive dependence of the polarization rate on the ionization potential we can neglect plasma formation in host material. 

The contribution from the plasma is determined by the average displacement $\langle d\rangle(z,t)$ of the electron from the equilibrium position in the
parent "molecule", whereby by a "molecule" we denote an atom or a group of atoms of the solid-state material which can provide a single ionization event. Furthermore, it is determined by the relative ionization of the solid state $\rho(z,t)$, which is the ratio of the conduction-band electron density to the density of "molecules":

\begin{equation}
    P_{\mathrm{plasma}}(z,\omega)=-N_{\mathrm{mol}}e\hat{F}[\langle d\rangle(z,t)\rho(z,t)]
\end{equation}

Here $N_{\mathrm{mol}}$ is the concentration of the molecules and $e=1.6\times 10^{-19}$ is the electron charge. 
The above expression would be valid in a homogeneous medium; however, as it refers to a polarization which occurs inside of NPs, in contrast to averaged macroscopic polarization, in the case of effective-medium theory it has to be additionally multiplied by $x(\omega)$. For the origin of this factor and further details see Ref. \cite{sb}.

The dynamics of the quantity
$\langle d\rangle(z,t)\rho(z,t)$ is given by \cite{h_highpower}

\begin{equation}
  \frac{\partial(\langle d\rangle(z,t)\rho(z,t))}{\partial t}=\langle v\rangle(z,t)+x_0\Gamma(t),
\end{equation}

where $\langle v\rangle$ is the average velocity of electrons and $x_0\simeq -I_p/eE(t)$ is the initial displacement of the electron immediately after the ionization event, $I_p$ being the bandgap. It can be shown that the second term describes the energy loss of the pulse due to the photoionization. The dynamics of the 
$\langle v\rangle(z,t)\rho(z,t)$ is given by second Newton's law as

\begin{equation}
  \frac{\partial(\langle d\rangle(z,t)\rho(z,t))}{\partial t}=-\frac{eE(z,t)}{m_e}\rho,  
\end{equation}

where $m_e$ is the effective electron mass near the bottom of the conduction band. Here we neglect the initial displacement and velocity of electron just after the ionization.

The dynamics of the relative plasma density $\rho$ is given by

\begin{equation}
    \frac{\partial\rho}{\partial t}=\Gamma(\hat{F}^{-1}[x(\omega)E(z,\omega)]),
\end{equation}

where $x(\omega)E(z,\omega)$ is the local field inside of inclusions which determines the 
photoionization rate $\Gamma$.

\begin{figure}[!ht]
\center{\includegraphics[width=0.45\textwidth]{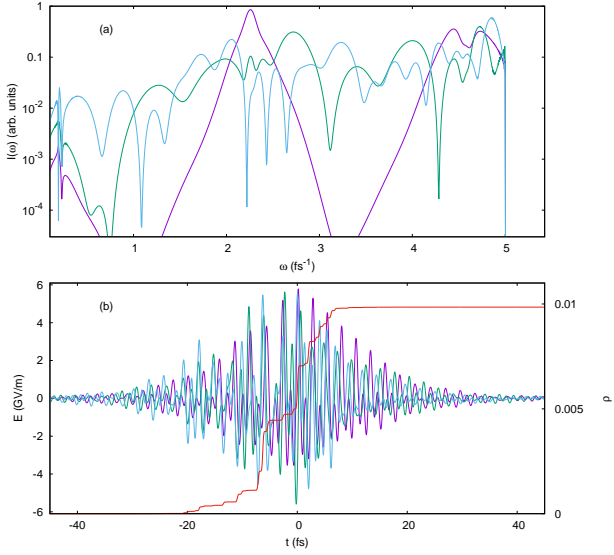}}
\caption{\label{main}Dependence of the spectra (a) and the electric field (b) on the propagation length. 15-fs pulses at 2.26 fs$^{-1}$ and 4.58 fs$^{-1}$ are considered, with  intensity of 1 TW/cm$^2$ and propagation length of 0.75 $\mu$m (magenta curves), 2.25 $\mu$m (green curves), and 6.75 $\mu$m (bluecurves). In (b), additionally the relative plasma density is shown for propagation length of 6.75 $\mu$m. A composite of ZnO inclusions with $f=0.03$ in fused silica is considered.}
\end{figure}

\subsection{Ionization rate}Depending on the relation between the frequency of pump light and the ionization potential of inclusions,
we consider two models for the ionization rate. For the case when the energy of pump photons is much smaller than the ionization potential, the photoionization occurs either in a multiphoton regime or in a tunneling regime, as determined by intensity and Keldysh parameter. Here we utilize so-called Yudin-Ivanov model \cite{yi}, which provides a formalism for both of these regimes in a unified way. This model was initially developed for isolated atoms; its use for solid state is justified in a case a negligible anharmonicity of the bands in the center of the Brillouin zone.

The cycle-resolved ionization rate $\Gamma$ is given (in atomic units, that is, with frequency $\omega$, time $t$ and field $\mathcal E$ measured in the corresponding Hartree units $\omega_a=0.26$ rad/as, $t_a=24.2$ as, $x_a=0.0529$ nm, and $\mathcal E_a=514.2$ V/nm) by

  \begin{eqnarray}
    \label{eq:YI}
    \Gamma(z,t)&=&\frac{\pi}{\tau_{T}}\exp\left(-\sigma_{0}\frac{
      \langle 2 \mathcal E(z,t)^{2}\rangle}{\omega^{3}}\right)\left[\frac{2\kappa^{3}}{
       \sqrt{\langle 2\mathcal E(z,t)^{2}\rangle}}\right]^{2Z/\kappa}\nonumber\\&\times &\exp\left[-\frac{ \mathcal E(z,t)^{2}}{2\omega^{3}}\sigma_{1}\right].
  \end{eqnarray}

Here $\tau_T=\kappa/E(z,t)$, $\kappa=\sqrt{I_p/(\hbar\omega_a)}$, $\sigma_{0}=\frac12(\gamma^{2}+\frac12)\ln
C-\frac12\gamma\sqrt{1+\gamma^{2}}$, $\gamma=\omega \tau_T$, $Z$ is the effective atomic charge, 
$C=1+2\gamma\sqrt{1+\gamma^{2}}+2\gamma^{2}$, and
$\sigma_{1}=\ln C$-$2\gamma/\sqrt{1+\gamma^{2}}$. The quantity $\langle E(z,t)^{2}\rangle$ is the 
averaged value of the squared electric field over few past periods (5 fs is assumed in this work).

The Yudin-Ivanov model was initially derived for gases; its applicability for solid state, while generally justified for materials with tight binding, is not strictly established. We have benchmarked Yudin-Ivanov model by comparing it to the numerical solution of the time-dependent Schrodinger equation in single active electron approximation  \cite{ab}. In this approach the empirical pseudo-potential method was used for describing the electron band structure of ZnO  \cite{fan}. We have found that the difference of the ionization rate does not typically exceed one order of magnitude. This difference is, in fact, not very significant: because
of the threshold-like behavior of the ionization rate, it leads to only a slight shift of the intensity at which a strong plasma generation is reached.

For the special case when the energy of pump photons is around two ionization potentials, it is preferable to use the
two-photon formalism \cite{chapter} and write the cycle-resolved ionization rate $\Gamma$ (in SI units) as

\begin{equation}
    \Gamma(z,t)=\frac{2e^4x_a^4\nu}{\hbar^4\omega_0^2[(2\omega_0-I_p/\hbar)^2+\nu^2}\langle  E(z,t)^{2}\rangle E(z,t)^{2},
\end{equation}

where $\nu$ is the relaxation constant of the two-photon transition.

\subsection{Contribution by excitons}Finally, we include the nonlinear polarization due to excitons into treatment. We consider multiple excitonic levels and utilize the standard Bloch equations for the description of the ionization. The dynamics of the density matrix $\rho_e$ is given by (see e.g. \cite{chapter})

\begin{equation}
  i\hbar\frac{\partial \rho_e}{\partial t}=[H,\rho_e],
\end{equation}

where $H=H_0+H_{\mathrm{int}}$, $H_0$ is the Hamiltonian of the system in the absence of excitation, $H_{\mathrm{int}}$ is the interaction  Hamiltonian, which components $H_{ij}$ are related with the corresponding 
dipole transition moments $eM_{ij}$:

\begin{equation}
H_{ij}=eM_{ij}\hat{F}^{-1}[E(z,\omega)x(\omega)].
\end{equation}

In addition, polarization decay (decay of the off-diagonal elements of $\rho_e$) with decay time $T_2$ and decay of the population to 
the ground state with decay time $T_1$ are included. In order to avoid numerical instabilities, the normalization of the density matrix $\rho$ is performed each few steps in time, by a) enforcing $0\le\rho_{e,ii}\le 1$, b) enforcing $Tr(\rho_e)=1$, and c) adjusting the non-diagonal elements which exceed the maximum possible value determined by the corresponding level populations.  

The excitonic polarization is then defined in a standard way as

\begin{equation}
    P_{\mathrm{exc}}(z,\omega)=x(\omega)\hat{F}[f\mathrm{Tr}(\rho_eM)].
\end{equation}

We solve the propagation equation by an extended split-step method, whereby each of the contributions to the polarization is treated subsequently, which allows to reduce the accumulation of numerical error. Nonlinear steps are performed using the Runge-Kutta approach, the order of which can be selected between 1,2, and 4. Fixed step of the grid both in time and in the propagation coordinate is used. The appearance of numerical artifacts during the propagation is monitored by tracing the total pulse energy as well as the total energy absorbed at the boundaries of the numerical time window. 

\section{Numerical results and discussion}

In order to exemplify the above model and functioning of SOLPIC, we present in this section a simulation of THz generation. We consider a composite of ZnO inclusions in SiO$_2$ matrix. Phenomenological Sellmeyer-type expressions were used to describe dispersion on ZnO \cite{disp_ZnO} and SiO$_2$ \cite{dispSiO2}. Similarly, experimental data on second-order \cite{sec_ZnO} and third-order \cite{thi_ZnO_1,thi_ZnO_1} susceptibility of bulk ZnO and third-order susceptibility of SiO$_2$ \cite{thi_SiO2} were used. We estimated the value of $T_2$ as 50 fs from Ref. \cite{T2} and used $T_1=2T_2$. For ZnO, the typical exciton size is larger than interatomic distance, meaning that we are dealing with Wannier-Mott type of excitons. In a case of sufficiently small inclusions, the exciton is bounded by the inclusion boundaries, therefore its wavefunctions (as well as energy levels and dipole momenta) are better described, instead of hydrogen-like potential, by a constant potential inside a sphere \cite{excit} with a step on its boundary. We have taken into account 5 lowest excitonic levels, and typical values of the off-diagonal dipole momenta, as calculated by this approach, are around 3$\times$10$^{-28}$ C$\cdot$m, for same-size NPs with a radius of 2.5 nm which are considered here and hereafter. For the permanent dipole momenta of ZnO, we have adopted a typical value of 6.66$\times$10$^{-30}$ C$\cdot$m per ZnO molecule, which was used to define the on-diagonal elements of the dipole matrix. We used the ionization potential of 3.37 eV equal to the bandgap of ZnO to characterize the transition from valence band to conduction band, and all the presented numerical results correspond to the conditions below the damage threshold of ZnO \cite{thres}.

The evolution of the field profile and spectra with propagation is illustrated in Fig. 1, for two-color pulsed excitation with pump pulses around 800 nm and 400 nm, for conditions given in the caption. In Fig. 1(a) one can see that initial stages of the propagation are characterized by self-phase modulation with typical spectral side lobes. At later stages, spectrum becomes irregular and transform into a supercontinuum extending up to the absorption edge given by the bandgap. The evolution of the temporal profile, shown in Fig. 1(b), shows gradual reduction of the energy of electric field, as well as significantly irregular envelope for longer propagation. This reduction of the maximum field determines the saturation of the THz generation efficiency and is caused both by strong group-velocity dispersion for broad spectrum and energy absorption due to transition to conduction band. One can see from the red curve in Fig. 1(b) that relative plasma density reaches values of roughly 0.01 after the pulse, which is sufficient to induce significant energy absorption.

\begin{figure}[!ht]
\center{\includegraphics[width=0.4\textwidth]{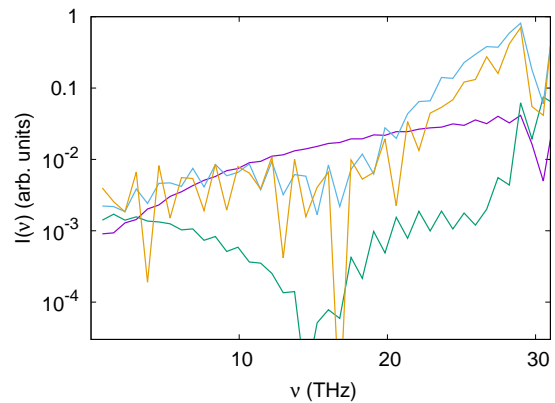}}
\caption{\label{thz}Dependence of the spectra in the THz range on the propagation distance. We consider 1-TW/cm$^2$, 15-fs pump pulses at 2.26 fs$^{-1}$ and 4.58 fs$^{-1}$, in a composite of ZnO NPs with filling fraction of $f=0.03$ in a fused-silica matrix, after propagation length of 5 $\mu$m (magenta curve), 15 $\mu$m (green curve), 45 $\mu$m (blue curve), and 50 $\mu$m (yellow curve).}
\end{figure}

In Fig. 2 the evolution of the spectrum in the THz range is shown. One can see that while the spectrum is flat at early stages of propagation, for larger propagation lengths the spectrum is localized around 28 THz, probably due to phase-matching effects with a phase-mismatch length of 10.5 $\mu$m for the four-wave mixing between the two photons at 2.26 fs$^{-1}$, one photon at 4.58 fs$^{-1}$, and a THz photon. Losses around 15 THz and below can also contribute to saturation of generation. After 45 $\mu$m propagation length, the efficiency of the generation reaches 3.05\%, which is sufficiently high for practical applications.

\begin{figure}[!ht]
\center{\includegraphics[width=0.45\textwidth]{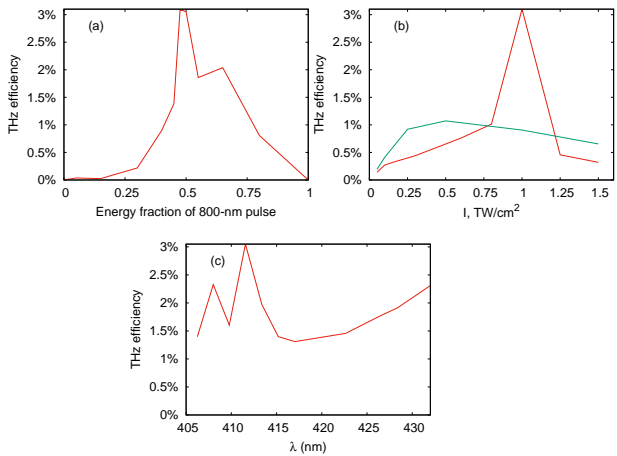}}
\caption{\label{eff_opt}Dependence of the THz generation efficiency on energy fraction of the 800-nm pulse (a), intensity of each of the pump pulses (b), and the wavelength of the second-harmonic pulse (c). A composite of ZnO NPs with $f=0.03$ in fused silica is considered. In (a), 15-fs pulses at 2.26 fs$^{-1}$ and 4.58 fs$^{-1}$ are considered, with total intensity of 2 TW/cm$^2$ and propagation length of 50 $\mu$m. In (b) we consider 15-fs (red curve) and 150-fs (green curve) pulses at 2.26 fs$^{-1}$ and 4.58 fs$^{-1}$. In (c), 1 TW/cm$^2$, 15-fs pulses are considered, with IR pulse frequency of 2.26 fs$^{-1}$ and propagation length of 50 $\mu$m.}
\end{figure}

In order to determine the optimum conditions of THz generation, in Fig. 3 we plot the dependence of the generation efficiency on the distribution of energy between the 830-nm pulse and 412-nm pulse (a), intensity of pulses (b), and wavelength of the short-wavelength pulse (c). One can see that the efficiency of THz generation is non-zero but very small for the cases when only one of the pulses is present (energy fraction of 0 or 1). This indicates that the optical rectification based the second-order susceptibility of ZnO cannot efficiently generate THz for the considered conditions, and that the dominant contribution comes from the third-order susceptibility of ZnO NPs, third-order susceptibility of SiO$_2$ being comparatively weak. In an ideal case without pump pulses modification, the efficiency of the THz generation is proportional to $E_{830}^2(E_{\mathrm{tot}}-E_{830})$, where $E_{830}$ is the energy of the pulse at 830 nm and $E_{\mathrm{tot}}=E_{830}+E_{412}$ is the total energy of the pulses. The maximum efficiency is then reached at $E_{830}/E_{\mathrm{tot}}=1/3$, however, as shown in Fig. 3(a), maximum numerical efficiency is achieved for $E_{830}/E_{\mathrm{tot}}=0.5$. This could be due to strong  SPM-induced spectral spreading of high-frequency pulse during the propagation, which needs to be compensated by relatively higher value of $E_{412}$. In Fig. 3(b), the dependence of the efficiency on the pulse intensity is shown, exhibiting saturation and decrease after a certain intensity as well as lower efficiencies for longer pulses. We attribute these features to detrimental contribution of the accumulated plasma, which grows with intensity and pulse duration [cf. Fig. 4(a)]. In Fig. 3(c), the dependence of the efficiency on the wavelength of the short-wavelength pulse exhibits several maxima. Note that while one  might expect an optimum THz generation for 415 nm, which would correspond to generation of frequencies near zero, our simulation in fact predict a minimum around this value,  determined most probably by phase mismatch and losses below 15 THz.

\begin{figure}[!t]
\center{\includegraphics[width=0.45\textwidth]{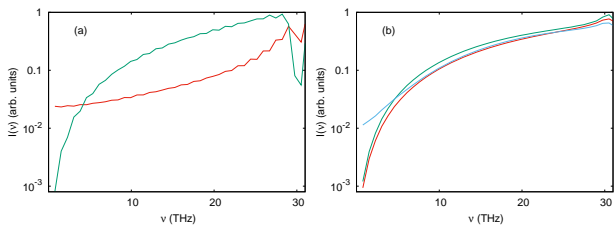}}
\caption{\label{onoff}Spectra in the THz range with (a) plasma contribution on (red) and off (green)  and (b) exciton contribution on (red) and off (green), as well including both exciton contribution and permanent dipole moment (blue). We consider 1-TW/cm$^2$, 15-fs pump pulses at 2.26 fs$^{-1}$ and 4.58 fs$^{-1}$, in a composite of ZnO inclusions with filling fraction of $f=0.03$ in a fused-silica matrix, after propagation length of 10 $\mu$m (a) and 0.75 $\mu$m (b).}
\end{figure}

Finally, in order to access the role of plasma and excitons in the THz generation in composites, in Fig. 4 we compare the spectra for plasma contribution (a) and exciton contribution (b) switched on/off. One can see that the plasma contribution is significant, both due to contribution to refractive index and due to losses, and absence of plasma contribution leads to a notable (more than twofold) increase of the efficiency. On the other hand, from Fig. 4(b) one can see that exciton polarization do not provide a strong contribution to the efficiency for the considered parameters. 
Also, additionally including the permanent dipole momenta, described in the model above, does not significantly increase the efficiency of THz generation, as indicated by the blue curve in Fig. 4(b) which is close to the red and green curves. We note, however, that this conclusion is of limited generality; for other parameters of the medium excitons can provide the key mechanism of THz generation (see e.g. \cite{exc_imp1,exc_imp2} and references therein).

\section{Conclusion}

In this paper we have established a comprehensive numerical model for the simulation of light propagation in composites, including all the relevant physical effects for a broad range of parameters, such as linear dispersion of the composite, second- and third-order nonlinear effects, plasma contribution, excitons contribution and so on. The model was applied to simulate the generation of THz radiation in a ZnO-SiO$_2$ composite. We have performed optimization of the frequency conversion process, predicting an efficiency of 3.05\%. We show that simulations provide insights into the optimization, such as the power distribution between the pump pulses, which would not be accessible intuitively. We hope that the numerical model and the corresponding software solution, which we make available for the community, will contribute to the capacity of the simulations in the area of nonlinear optics.

\begin{acknowledgments}
Authors acknowledge financial support from European Union project H2020-MSCA-RISE-2018-823897 "Atlantic".
I.B. thanks Cluster of Excellence PhoenixD (EXC 2122, project ID 390833453) for financial support. 
Support from the BNSF under Contract No. KP-06-COST/7 is acknowledged (T.A.) 

\end{acknowledgments}

\end{document}